\pdfoutput=1

\documentclass[prl,twocolumn,superscriptaddress]{revtex4}
\usepackage[]{graphicx}
\usepackage{amssymb,amsmath,multirow}

\usepackage{amsfonts,amssymb,amsmath}
\usepackage[]{graphics,graphicx,epsfig}
\usepackage{amsthm}
\usepackage[colorlinks=true,citecolor=blue,urlcolor=blue]{hyperref}

\begin{document}


\title{Fundamental Monogamy Relation between Contextuality and Nonlocality}



\author{Pawe\l\ Kurzy\'nski}
\email{cqtpkk@nus.edu.sg}
\affiliation{Centre for Quantum Technologies,
National University of Singapore, 3 Science Drive 2, 117543 Singapore,
Singapore}
\affiliation{Faculty of Physics, Adam Mickiewicz University,
Umultowska 85, 61-614 Pozna\'{n}, Poland}

\author{Ad\'an Cabello}
 \email{adan@us.es}
 \affiliation{Departamento de F\'{\i}sica Aplicada II, Universidad de Sevilla, E-41012 Sevilla, Spain}

\author{Dagomir Kaszlikowski}
\email{phykd@nus.edu.sg}
\affiliation{Centre for Quantum Technologies,
National University of Singapore, 3 Science Drive 2, 117543 Singapore,
Singapore}
\affiliation{Department of Physics,
National University of Singapore, 3 Science Drive 2, 117543 Singapore,
Singapore}


\begin{abstract}
We show that the no-disturbance principle imposes a tradeoff between locally contextual correlations violating the Klyachko-Can-Binicio\u{g}lu-Shumovski inequality and spatially separated correlations violating the Clauser-Horne-Shimony-Holt inequality. The violation of one inequality forbids the violation of the other. We also obtain the corresponding monogamy relation imposed by quantum theory for a qutrit-qubit system. Our results show the existence of fundamental monogamy relations between contextuality and nonlocality that suggest that entanglement might be a particular form of a more fundamental resource.
\end{abstract}

\maketitle


{\em Introduction.---}Since its inception, quantum theory (QT) has radically altered our understanding of nature. The pioneering works of Kochen and Specker (KS) \cite{Specker60,KS67} and Bell \cite{Bell66,Bell64}, and the subsequent experiments \cite{FC72,ADR82} demonstrated that nature denies the possibility of noncontextual and local hidden variables.

Theoretical proofs of the impossibility of hidden variables fall into two seemingly distinct classes. In KS-like proofs, there is a {\it single observer} that performs measurements on a physical system, whereas in Bell proofs there are {\it two spatially separated observers}, customarily called Alice and Bob, each performing some measurements on their respective physical systems. Both classes have a common trait: both ask whether a joint probability distribution for all the measurements exists and in both cases this question can be recast as whether or not some set of correlation inequalities is violated \cite{Fine,AQBTC12}.

For the KS scenario, the simplest inequality violated by QT is the Klyachko-Can-Binicio\u{g}lu-Shumovski (KCBS) inequality \cite{KCBS08}. Its quantum violation requires, at least, a single three-dimensional quantum system (qutrit) and has it been experimentally observed recently \cite{AACB13,Lapkiewicz}. For the Bell scenario, the simplest inequality violated by QT is the Clauser-Horne-Shimony-Holt (CHSH) inequality \cite{CHSH69}. Its quantum violation requires a minimal system of two spatially separated two-dimensional quantum systems (qubits) and it has been experimentally observed numerous times since the seminal experiments in Refs.~\cite{FC72,ADR82}.

In this Letter we qualitatively and quantitatively study treadoffs between these two fundamental inequalities imposed by the principle of no-disturbance (ND) as well as by QT.

The ND principle is a generalization of the no-signaling principle that refers to compatible observables instead of spacelike separated observables \cite{Gleason,CSW,Amselem12,Monogamy2}. It states that, for any three observables $A$, $B$, and $C$ such that $A$ and $B$ are compatible, and $A$ and $C$ are compatible, the probabilities of outcomes of $A$ do not depend on whether $A$ was measured with $B$ or with $C$,
\begin{equation}
 p(a)=\sum_{b} p(a,b) = \sum_c p(a,c).
\end{equation}
QT satisfies the ND principle, but there are theories that satisfy the ND principle and violate noncontextuality and Bell inequalities more than QT.

We first show that, in an experiment to test KCBS correlations within a single system and CHSH correlations between this system and another system, the violation of one inequality forbids the violation of the other, defining a so-called monogamy (in analogy to Refs.~\cite{OV06,TV06}). This monogamy is implied by the ND principle and can be derived without any reference to QT. We also obtain the quantum version of this monogamy and show that the tradeoff between the violations of the KCBS and CHSH inequalities is more stringent in QT than the one resulting from the ND principle.

\begin{figure}[t]
\begin{center}
\includegraphics[width=0.76\columnwidth]{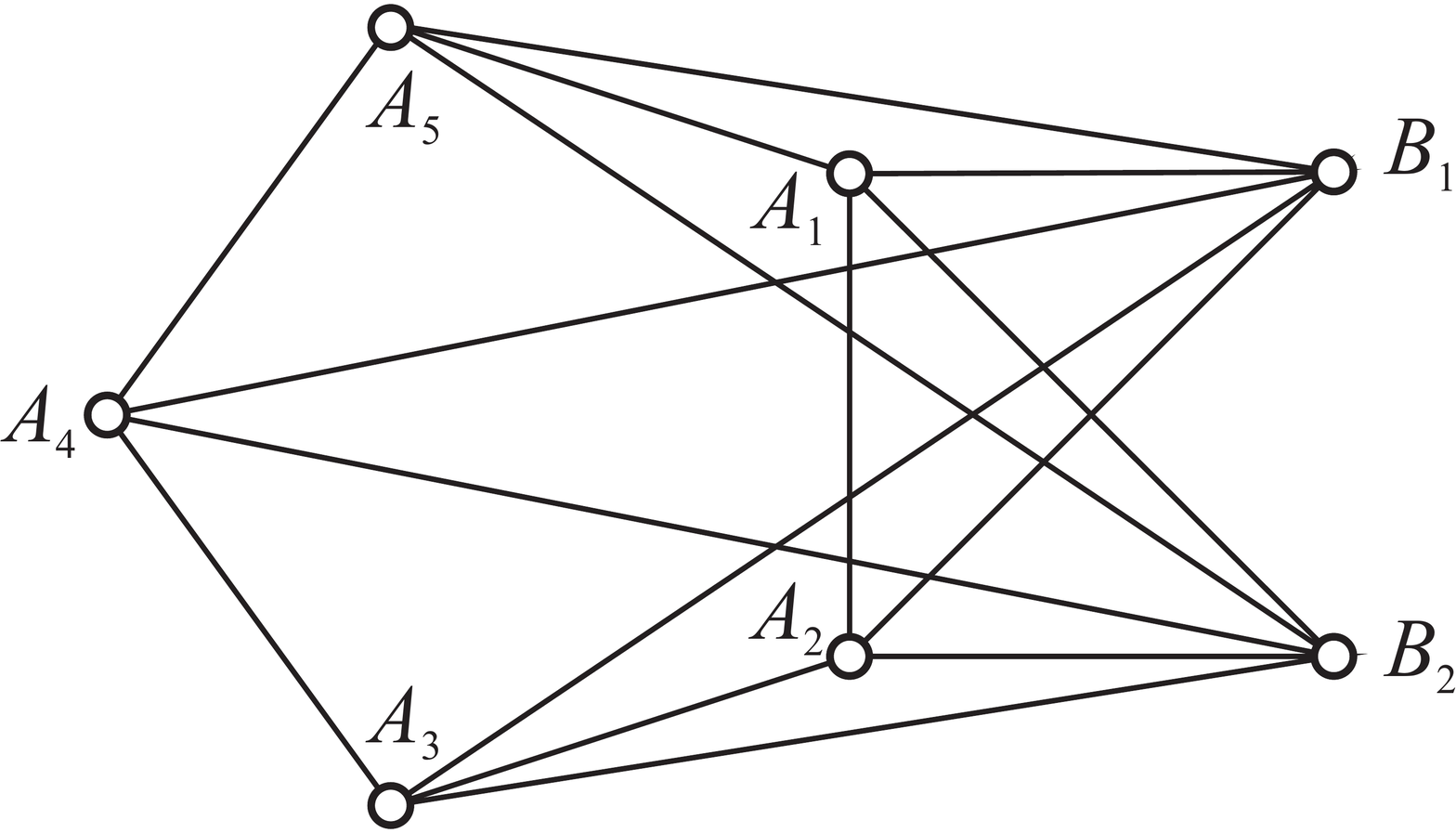}
\caption{\small Compatibility graph corresponding to the measurements in the scenario for the monogamy relation between locally contextual and nonlocal correlations. $A_1,\ldots,A_5$ are five cyclically compatible measurements on Alice's system, and $B_1$ and $B_2$ are two incompatible measurements on Bob's system. Vertices represent measurements and adjacent vertices represent pairwise compatible measurements.}
\label{Fig1}
\end{center}
\end{figure}

{\em No-disturbance monogamy between the KCBS and CHSH inequalities.---}Consider the following scenario: Alice and Bob share pairs of correlated systems. Alice can perform five measurements $\{A_1,\ldots,A_5\}$ on her system. Each measurement has two outcomes $\pm 1$ and the measurements $A_i$ and $A_{i+1}$ (with the sum modulo 5) are compatible. For each pair of systems, Alice randomly chooses two compatible measurements $A_i$ and $A_{i+1}$, and Bob randomly chooses only one of two incompatible measurements $B_1$ or $B_2$, each with outcomes $\pm 1$. The compatibility relations among the seven measurements are illustrated in Fig.~\ref{Fig1}.


After many rounds of the experiment, Alice and Bob can evaluate the following correlations (mean values of products of outcomes):
\begin{equation}
\langle A_i A_{i+1}\rangle,~~\langle A_i B_j\rangle,~~\langle A_i A_{i+1} B_j\rangle,
\end{equation}
where $i=1,\ldots,5$ and $j=1,2$. These correlations can be used in two different tests. The first one is a test of the KCBS noncontextuality inequality on Alice's system, i.e.,
\begin{eqnarray}\label{KCBS}
\kappa_A &=& \langle A_1 A_2\rangle + \langle A_2 A_3\rangle + \langle A_3 A_4\rangle + \langle A_4 A_5\rangle \nonumber \\ &+& \langle A_5 A_1\rangle \stackrel{\mbox{\tiny{ NCHV}}}{\geq} -3.
\end{eqnarray}
The second one is a test of the CHSH Bell inequality between Alice's and Bob's subsystems, i.e.,
\begin{equation}\label{CHSH}
\beta_{AB}=\langle A_{i+1} B_1 \rangle + \langle A_{i+1} B_2 \rangle + \langle A_{i-1} B_1 \rangle - \langle A_{i-1} B_2 \rangle \stackrel{\mbox{\tiny{ LHV}}}{\geq} -2,
\end{equation}
where $A_{i+1}$ and $A_{i-1}$ can be any two incompatible measurements from Alice's set.

Both the KCBS and CHSH inequalities are tight in the sense that the violation of each inequality implies that the corresponding correlations cannot be described either via a noncontextual hidden variable (NCHV) model or a local hidden variable (LHV) model. The lack of violation implies the existence of such a model. The existence of a noncontextual or local hidden variable model is equivalent to the existence of a joint probability distribution over all the observables involved in the inequality \cite{Fine,AQBTC12,LSW}.

An important property of both inequalities is that the classical bounds of $-3$ and $-2$ result from the noncontextuality assumption in both cases. Also in both cases, these bounds can be violated due to the lack of a joint probability distribution. Nevertheless, the maximal violations can also be bounded. These contextual bounds may result from various principles \cite{IC,Cabello13,Yan13}. One of such principles is the no-disturbance (ND) principle.

The ND bound of the KCBS inequality is $-5$ and for the CHSH inequality the ND bound is $-4$. Although these bounds are the same as the algebraic bounds of both inequalities, the ND principle leads to a nontrivial tradeoff between their violations.

Let us prove the monogamy of the inequalities (\ref{KCBS}) and (\ref{CHSH}) using the techniques of Refs.~\cite{Monogamy1,Monogamy2}. Since Bell inequalities are also noncontextuality inequalities \cite{G10}, one can sum both inequalities and produce a new noncontextuality inequality and split the terms into two new groups $C_1^{(i)}$ and $C_2^{(i)}$
\begin{equation}
 C_1^{(i)}+C_2^{(i)} \stackrel{\mbox{\tiny{ NCHV}}}{\geq} -5,
\end{equation}
where
\begin{subequations}
\begin{align}
 C_1^{(i)} =& \langle A_{i+1} B_1\rangle + \langle A_{i+1} A_{i+2}\rangle + \langle A_{i+2} A_{i-2}\rangle \nonumber \\ &+ \langle A_{i-2} A_{i-1}\rangle + \langle A_{i-1} B_1\rangle, \label{uno}\\
 C_2^{(i)} =& \langle A_{i+1} A_{i}\rangle + \langle A_{i-1} A_{i}\rangle + \langle A_{i+1} B_2\rangle - \langle A_{i-1} B_2\rangle. \label{dos}
\end{align}
\end{subequations}
Note that $C_1^{(i)}$ and $C_2^{(i)}$ have the form of the KCBS and CHSH expressions, respectively.

For any theory satisfying the ND principle, the lower bounds for $C_1^{(i)}$ and $C_2^{(i)}$ are the same as those for NCHV theories, namely,
\begin{subequations}
\begin{align}
 & C_1^{(i)} \stackrel{\mbox{\tiny{ ND}}}{\geq} -3, \label{c1} \\
 & C_2^{(i)} \stackrel{\mbox{\tiny{ ND}}}{\geq} -2. \label{c2}
\end{align}
\end{subequations}
This is because $B_1$ and $B_2$ are compatible with all measurements $A_i$ and the ND principle allows one to construct joint probability distributions recovering the measurements statistics involved in $C_1^{(i)}$ and $C_2^{(i)}$. Let us simplify the notation by using $p(a_i)=p(A_i=a_i)$. The experimental probabilities $p(a_i,a_{i+1},b_j)$ allow us to calculate the marginal probabilities $p(a_i,a_{i+1})$, $p(a_i,b_j)$, $p(a_i)$, and $p(b_j)$. Following Ref.~\cite{Entropic}, one can construct a joint probability distribution from which $C_1^{(i)}$ can be derived, in the following way:
\begin{equation}\label{prob1}
\begin{split}
 & p(a_{i+1},a_{i+2},a_{i-1},a_{i-2},b_1) \\
 & =\frac{p(a_{i+1},a_{i+2},b_1)p(a_{i+2},a_{i-2},b_1)p(a_{i-1},a_{i-2},b_1)}{p(a_{i+2}b_1)p(a_{i-2}b_1)}.
\end{split}
\end{equation}
Similarly, the joint probability distribution from which $C_2^{(i)}$ can be derived is
\begin{equation}\label{prob2}
 p(a_{i-1},a_i,a_{i+1},b_2)=\frac{p(a_i,a_{i-1},b_2)p(a_i,a_{i+1},b_2)}{p(a_i,b_2)}.
\end{equation}
The above joint probability distributions recover all measurable marginals.

The ND principle is necessary for the validity of the above construction. For example, one may calculate $p(a_{i-1},a_i)$ from $p(a_{i-1},a_i,a_{i+1},b_2)$ in the following way:
\begin{equation}
\begin{split}
 &\sum_{a_{i+1},b_2}p(a_{i-1},a_i,a_{i+1},b_2) \\
 &=\sum_{b_2}\left(\sum_{a_{i+1}} \frac{p(a_i,a_{i-1},b_2)p(a_i,a_{i+1},b_2)}{p(a_i,b_2)}\right) \\
 &=\sum_{b_2} p(a_i,a_{i-1},b_2) = p(a_i,a_{i-1}).
\end{split}
\end{equation}
On the other hand, one may also calculate $p(a_{i+1},a_i)$ as
\begin{equation}
\begin{split}
 &\sum_{a_{i-1},b_2}p(a_{i-1},a_i,a_{i+1},b_2) \\
 &=\sum_{b_2}\left(\sum_{a_{i-1}} \frac{p(a_i,a_{i-1},b_2)p(a_i,a_{i+1},b_2)}{p(a_i,b_2)}\right) \\
 &=\sum_{b_2} p(a_i,a_{i+1},b_2) = p(a_i,a_{i+1}).
\end{split}
\end{equation}
In both derivations we assumed
\begin{equation}
\sum_{a_{i-1}} p(a_i,a_{i-1},b_2) = \sum_{a_{i+1}} p(a_i,a_{i+1},b_2) = p(a_i,b_2),
\end{equation}
which is exactly the ND principle.

Note that the probabilities on the left hand sides of Eqs. (\ref{prob1}) and (\ref{prob2}) may not be directly defined within the ND theory (like quantum theory); i.e., the theory may not allow for a direct evaluation of the joint probability for all measurements. However, the constructions on the right hand sides take into account only the probabilities that are measurable in the laboratory. These probabilities are defined within ND theories, since these theories aim to explain the experimental data. Moreover, the above constructions recover all measurable marginals compliant with ND theories.

The existence of a joint probability distribution for $C_1^{(i)}$ and $C_2^{(i)}$ guarantees that the inequalities (\ref{c1}) and (\ref{c2}) are satisfied \cite{Fine,LSW}, and that their sum is always bounded from below by $-5$ in any ND theory. This, in turn, implies that in any ND theory there is a monogamy relation between the KCBS and CHSH inequalities; i.e., only one of them can be violated:
\begin{equation}
 \label{monogamy}
 \beta_{AB} + \kappa_{A} \stackrel{\mbox{\tiny{ ND}}}{\geq} -5.
\end{equation}

The scenario discussed before can be easily extended to the case in which Bob, instead of two incompatible measurements $B_1$ and $B_2$, has five cyclically compatible measurements $A'_j$ ($j=1,\ldots,5$) and he also tries to violate the KCBS inequality on his system.
On the other hand, there is no reason for any monogamy between the KCBS tests of Alice and Bob, since they can always prepare their local systems independently in such a way that local measurements will violate the KCBS inequality.


{\em Quantum monogamy between local contextuality and nonlocality.---}The monogamy relation (\ref{monogamy}) holds in any theory satisfying the ND principe such as QT. A natural question is whether QT imposes an additional monogamy relation between {\em quantum} contextual and nonlocal correlations, similar to the quantum monogamy of nonlocality found in Ref.~\cite{TV06}.

To study this, let us consider a quantum mechanical implementation of the scenario described above in which Alice has five measurements $A_i$ ($i=1,\ldots,5$) and Bob two measurements $B_1$ and $B_2$. Alice tries to violate the KCBS inequality on her system and, in addition, Alice and Bob try to violate the CHSH inequality using Alice's incompatible measurements $A_1$ and $A_4$, namely,
\begin{equation}
 \label{CHSH2}
 \langle A_{1} B_1 \rangle + \langle A_{1} B_2 \rangle + \langle A_{4} B_1 \rangle - \langle A_{4} B_2 \rangle \stackrel{\mbox{\tiny{ LHV}}}{\geq} -2.
\end{equation}
We assume that Alice's system is a qutrit and Bob's system is a qubit. The corresponding basis states are $\{|0\rangle\,|1\rangle,|2\rangle\}$ and $\{|0\rangle,|1\rangle\}$, respectively.

We also assume that Alice's measurements are of the form
\begin{equation}
 A_i=2|v_i\rangle\langle v_i| -\openone,
\end{equation}
where $\langle v_i|v_{i+1}\rangle=0$ (with the sum modulo 5) and they are chosen to maximize the violation of the KCBS inequality. In particular, we assume
\begin{equation}
|v_i\rangle={\cal N} \left[ \cos\left(\frac{4\pi i}{5}\right)|0\rangle + \sin\left(\frac{4\pi i}{5}\right)|1\rangle+ \sqrt{\cos\left(\frac{\pi}{5}\right)}|2\rangle \right],
\label{vec}
\end{equation}
where ${\cal N}$ is a normalization constant. On the other hand, for the CHSH scenario we choose Bob's observables to be two Pauli operators $B_1=Z$ and $B_2=X$.

The form of vectors (\ref{vec}) makes the KCBS operator diagonal in the computational basis. The eigenvalues of this operator are highly degenerated
\begin{equation}
\begin{split}
&\lambda_1=\lambda_2=\lambda_3=\lambda_4=-5+2\sqrt{5}, \\
&\lambda_5=\lambda_6=5-4\sqrt{5}
\end{split}
\end{equation}
and correspond to eigenvectors
\begin{equation}
\begin{split}
&|\lambda_1\rangle=|00\rangle,\; |\lambda_2\rangle=|01\rangle,\;|\lambda_3\rangle=|10\rangle,\;|\lambda_4\rangle=|11\rangle, \\
&|\lambda_5\rangle=|20\rangle,\;|\lambda_6\rangle=|21\rangle.
\end{split}
\end{equation}

On the other hand, the CHSH operator can be written as a direct sum $M \oplus -M$, where
\begin{equation}
M=\begin{pmatrix} 1-\frac{1}{\sqrt{5}} & -\sqrt{2-\frac{2}{\sqrt{5}}} & \sqrt{\frac{4}{5}+\frac{4}{\sqrt{5}}} \\ -\sqrt{2-\frac{2}{\sqrt{5}}} & 1-\sqrt{5} & 2\sqrt{-1+\frac{3}{\sqrt{5}}} \\ \sqrt{\frac{4}{5}+\frac{4}{\sqrt{5}}} & 2\sqrt{-1+\frac{3}{\sqrt{5}}} & 2-\frac{4}{\sqrt{5}} \end{pmatrix}
\end{equation}
is represented in the basis $\{|01\rangle,|10\rangle,|21\rangle\}$. The second matrix $-M$ is the same matrix as $M$ (multiplied by $-1$); however, it is represented in the basis $\{|00\rangle,|11\rangle,|20\rangle\}$, respectively. The eigenvectors of $M$ are
\begin{subequations}
\begin{align}
 &|\lambda_1\rangle=\frac{1}{\sqrt{2\sqrt{5}}} |01\rangle + \frac{1}{\sqrt{2}}|10\rangle - \sqrt{\frac{1}{2}-\frac{1}{2\sqrt{5}}}|21\rangle, \\
  &|\lambda_2\rangle=\sqrt{1-\frac{1}{\sqrt{5}}} |01\rangle + \frac{1}{5^{1/4}} |21\rangle, \\
  &|\lambda_3\rangle=\frac{1}{\sqrt{2\sqrt{5}}} |01\rangle - \frac{1}{\sqrt{2}}|10\rangle - \sqrt{\frac{1}{2}-\frac{1}{2\sqrt{5}}}|21\rangle,
\end{align}
\end{subequations}
and the corresponding eigenvalues are
\begin{equation}
\lambda_1 \approx -2.808,~\lambda_2 = 2,~\lambda_3 \approx 0.336.
\end{equation}

The diagonal form of the KCBS operator and the separation of the CHSH operator into two symmetric parts reduce our monogamy problem to three real dimensions. Due to this symmetry, every quantum state has to produce a point that lies inside the region parametrized by only two real parameters corresponding to this three-dimensional space. It is important to notice that the KCBS operator can be written as $N \oplus N$, where $N=\text{diag}\{-5+2\sqrt{5},-5+2\sqrt{5},5-4\sqrt{5}\}$ is represented in the same basis as $M$. Therefore, it is enough to consider the monogamy problem between operators $M$ and $N$.

One can represent the quantum region corresponding to operators $M$ and $N$ using the following parametrization:
\begin{equation}
|\theta,\varphi\rangle=\cos\theta |a\rangle + \sin\theta\cos\varphi |b\rangle + \sin\theta\sin\varphi |c\rangle.
\end{equation}
The basis in the above formula is for convenience taken to be $|a\rangle=(0,0,1)^T$ (vector that maximizes $N$), $|b\rangle=(\alpha,\beta,0)^T$, $|c\rangle=(-\beta,\alpha,0)^T$, where $\alpha \approx 0.42$ and $\beta \approx 0.91$. The last two vectors are chosen to minimize and maximize $M$, respectively, for the minimal value of $N$. In other words, $|b\rangle$ and $|c\rangle$ are eigenvectors of the upper-left $2\times 2$ submatrix of $M$. The above parametrization gives
\begin{subequations}
\begin{align}
 \langle M\rangle_{\theta,\varphi} &=\gamma_1 \cos^2 \theta + (\gamma_2 + \gamma_3 \cos 2\varphi) \sin^2 \theta \\ & + \cos\theta \sin\theta (\gamma_4 \cos \varphi +\gamma_5 \sin\varphi), \nonumber \\
 \langle N \rangle_{\theta} &=-\sqrt{5}+(5-3\sqrt{5})\cos 2\theta,
\end{align}
\end{subequations}
where $\gamma_1 \approx 0.21$, $\gamma_2 \approx -0.34$, $\gamma_3 \approx -1.38$, $\gamma_4 \approx 3.47$, and $\gamma_5 \approx -1.94$.

The region corresponding to $-M$ and $N$ is given by the same formulas, i.e., $-\langle M\rangle_{\theta,\varphi}$ and $\langle N \rangle_{\theta}$. However, this time the computational basis $\{|01\rangle,|10\rangle,|21\rangle\}$ has to be changed to $\{|00\rangle,|11\rangle,|20\rangle\}$.

\begin{figure}[t]
\begin{center}
\includegraphics[width=1.00\columnwidth]{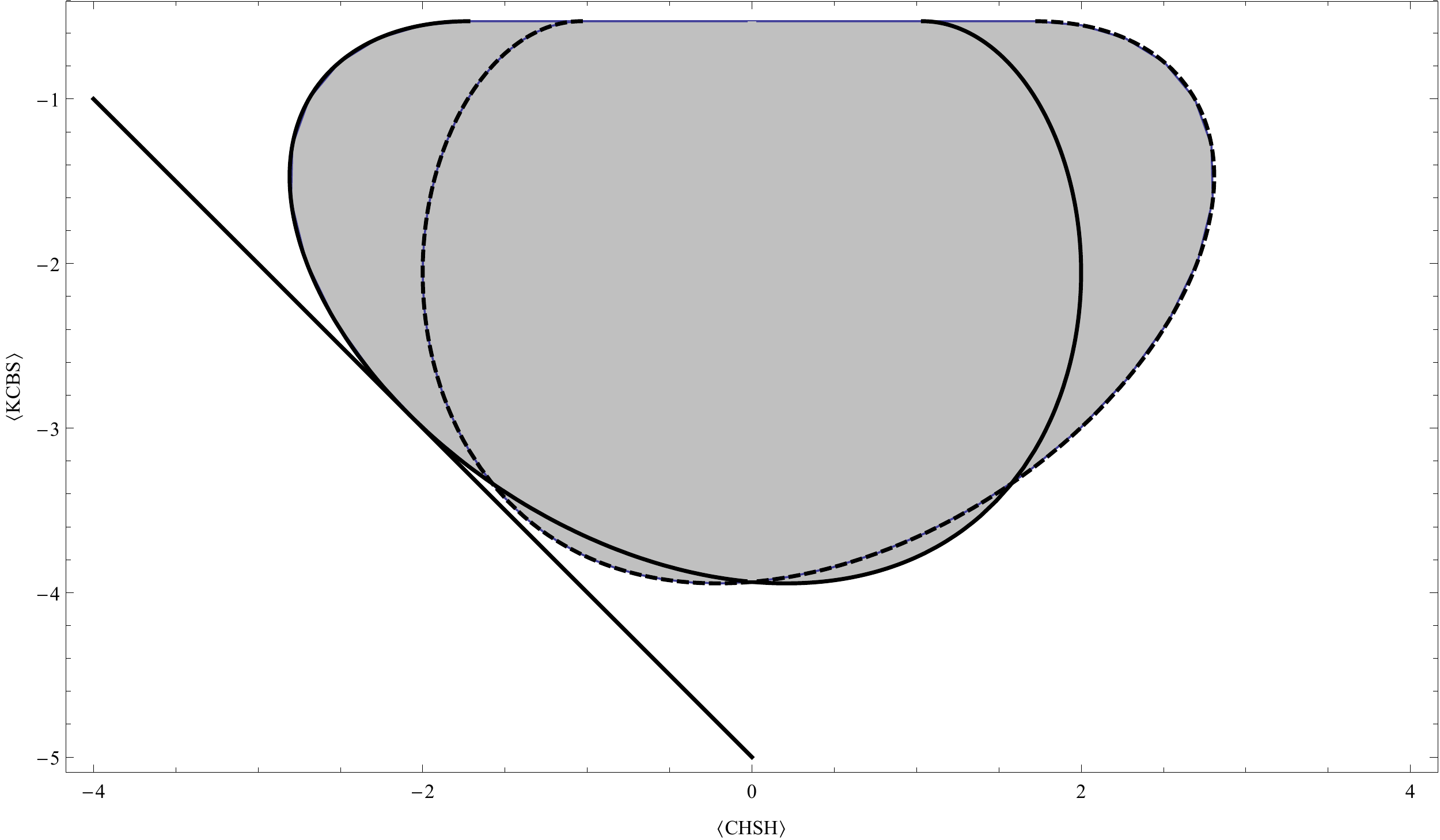}
\caption{\small Allowed average values of CHSH and KCBS operators. The region can be divided into two overlapping parts. The first part is spanned by vectors that are linear combinations of $\{|01\rangle,|10\rangle,|21\rangle\}$ and is bounded by the solid curve. The second region corresponds to the basis $\{|00\rangle,|11\rangle,|20\rangle\}$ and is bounded by the dashed curve. The solid line corresponds to the no-disturbance/no-signaling bound.}
\label{fig2}
\end{center}
\end{figure}

The boundaries of these regions can be analytically obtained via minimization and maximization of $\langle M\rangle_{\theta,\varphi}$ (or $-\langle M\rangle_{\theta,\varphi}$). This procedure reduces one of two parameters. In particular, one obtains
\begin{equation}
\tan{\theta}=\frac{\csc 2\varphi(\gamma_5 \cos\varphi+\gamma_4 \sin\varphi )}{2\gamma_3}.
\end{equation}
Both regions and their corresponding boundaries are represented in Fig.~\ref{fig2}.


Interestingly, the quantum boundary touches the no-disturbance and no-signaling boundary in a single point; however, unlike in the case of monogamy of two CHSH inequalities, this point does not correspond to classical bounds; i.e., it is not of the form $(\langle CHSH \rangle=-2,\langle KCBS \rangle=-3)$. Instead, this point is $(\langle CHSH \rangle\approx-2.08,\langle KCBS \rangle=-2.92)$. We speculate that the classical point can be achieved using different measurement settings and perhaps a larger system.

Finally, the two classes of (un-normalized) states recovering both quantum boundaries are
\begin{subequations}
\begin{align}
&|\psi_\varphi^+\rangle=f(\varphi)|01\rangle + g(\varphi)|10\rangle + |21\rangle,\\
&|\psi_\varphi^-\rangle=f(\varphi)|00\rangle + g(\varphi)|11\rangle +|20\rangle,
\end{align}
\end{subequations}
where
\begin{subequations}
\begin{align}
&f(\varphi)\approx -0.05 + 0.15 \cot \varphi -0.57\tan\varphi, \\
&g(\varphi)\approx 0.72 + 0.32 \cot \varphi +0.26\tan\varphi.
\end{align}
\end{subequations}
Due to the fact that the boundaries of both regions cross, one has to switch between the two classes in order to reproduce the total boundary of the quantum region.


{\em Conclusions.---}Previous works on contextuality and nonlocality monogamy relations had identified tradeoffs between the violation of either noncontextuality inequalities of the same type \cite{Monogamy2} or Bell inequalities of the same type \cite{Monogamy1,TV06,Monogamy3}. However, a fundamental question was whether similar relations exist between contextual correlations and nonlocal correlations. Here we have shown that this is the case. Specifically, we have shown that the ND principle imposes a fundamental monogamy relation between the local violation of the KCBS inequality, the simplest nocontextuality inequality violated by QT, and the nonlocal violation of the CHSH inequality, the simplest Bell inequality violated by QT. In addition, we have shown that QT imposes an even stronger restriction. This monogamy between contextuality and nonlocality can be experimentally observed in qutrit-qubit systems.

Although further exploration both theoretically and experimentally is needed, our results show the existence of fundamental monogamy relations between (local) contextuality and nonlocality and suggest that monogamy relations between different types of correlations might be ubiquitous in nature.

A final fundamental observation can be made. In QT, monogamy of nonlocality \cite{TV06} follows from monogamy of entanglement \cite{OV06}, a quantum resource that may be differently distributed among the parties \cite{CKW00}. A similar reasoning applied to the contextuality-nonlocality monogamy discussed in this Letter suggests the existence of a quantum resource of which entanglement is just a particular form. This follows from the fact that the resource needed to violate the KCBS inequality consumes the entanglement needed to violate the CHSH inequality and, at the same time, can be transformed into entanglement. This more general resource requires further investigation.


\begin{acknowledgments}
A.\ C. is supported by the Project No. FIS2011-29400 (MINECO, Spain) with FEDER funds and by grant ``The nature of information in sequential quantum measurements'' from the Foundational Questions Institute. P.\ K. and D.\ K. are supported by the National Research Foundation and Ministry of Education in Singapore and by grant ``Operational and information theoretic meaning of contextuality'' from the Foundational Questions Institute. D.\ K. is also supported by the Singapore Ministry of Education through the Academic Research Fund Tier 3 MOE2012-T3-1-009.
\end{acknowledgments}



\end{document}